\documentclass[twocolumn,showpacs,aps,floatfix]{revtex4}
\usepackage{amsmath}
\usepackage{graphicx}

\begin{document}

\title{Calculations of liquid helium and neon VUV emission spectra, self-absorption and scattering for a neutrino detector}
\author{I. M. Savukov}
\affiliation{Department of Physics, Princeton University,
Princeton, New Jersey 08544}
\date{\today}

\begin{abstract}
To evaluate the feasibility of the recently proposed detection
scheme of low energy neutrinos released from the Sun and
supernovae called CLEAN, Cryogenic Low Energy Astrophysics with
Noble Gases, which relies on the transparency of noble-gas
cryogenic liquids to VUV radiation produced by neutrinos, we
analyze theoretically VUV emission, self-absorption, and
scattering of liquid helium and neon, primary candidates for
CLEAN. Owing to strong repulsion of noble-gas atoms in the ground
states at the equilibrium distance of the relevant excited state,
the emission spectrum is substantially shifted from the absorption
spectrum, and in principle the absorption is expected very small,
allowing building large detectors. Our analysis, however, shows
that the self-absorption and Rayleigh scattering are comparable to
the size of the proposed detector.

Our theoretical emission spectra are found in agreement with
experimental observations although some deviation exists due to
binary-interaction approximation, and our ab initio Rayleigh
scattering lengths are found in agreement with other calculations
based on the extrapolation of experimental refraction indices. The
absorption process can result in either re-emission, which
conserves the number of photons but delays their escape from the
liquid, or in non-radiative quenching.
\end{abstract}
% pacs in sumbitted version are different
 \maketitle

\homepage{http://www.princeton.edu/~isavukov}

\affiliation{Department of Physics, Princeton University,
Princeton, New Jersey 08544}

Recently,~\citet{McKinsey} has proposed the detection of low
energy neutrinos released from the Sun and supernovae with the
scheme called CLEAN, Cryogenic Low Energy Astrophysics with Noble
Gases. The central element of a CLEAN detector is a 6-meter
diameter Dewar filled with liquid neon which is functioning both
as a passive shielding medium and an active self-shielding
detector. The advantage of liquid neon is high scintillation yield
and low radiactive background. Other noble-gas liquids can be used
as well, of particular interest is liquid helium. The
self-absorption and scattering of UV light produced by these
liquids are critical for determination of the size of the
detector, the interpretation of neutrino events, and exclusion of
background events. The emission of liquid helium and neon in UV
has already been studied~\cite{Stockton,Surko,Packard} and
scattering lengths of noble-gas liquids at scintillation
wavelengths has been estimated from extrapolations of refraction
indices~\cite{Seidel}; the self-absorption is yet to be measured.
Because there are many factors such as spectral reshaping and the
effects of impurities (for example, 0.2\% of N$_{2}$ impurities in
liquid helium reduces the emission intensity by 20-40\% in the
region 700-850\AA\ \cite{Surko}) that influence the interpretation
of the experimental results, the theoretical calculations of
emission, self-absorption, and scattering, which constitute the
subject of this letter, are important. Spectral properties of
noble-gas liquids are not well understood mostly due to
interdisciplinary nature (atomic, molecular, condensed-matter
physics) of the phenomenon as well as the scarcity of the data, so
this paper also has the aim of attracting the attention of
theorists of diverse expertise and motivating further UV
spectroscopic measurements.

The scintillation of liquid helium due to energetic $\alpha $-particle
stopped in the liquid was first studied by~\citet{Moss} and by~\citet{Kane}. %
\citet{Moss} observed significant decrease of the intensity below
$\lambda$-point, and attributed this to the change of the
oscillator strength of 2P-1S He transition due to the differences
in spatial distribution of nearest neighbors in the superfluid and
the normal fluid.
The emission spectrum of liquid helium was also studied experimentally by~%
\citet{Stockton} who obtained and compared UV spectra of
electron-bombarded superfluid helium at 1.4 and 2.1 K with those
of high-pressure helium discharge lamp in the range of 1000-600
\AA . The spectra were very similar, but the liquid spectrum had
an earlier cut-off at about 600 \AA\ due to stronger absorption.
\citet{Keto} and~\citet{Hill} showed that the ionization of the
helium results in formation of He$_{2}^{\ast }(A^{1}\Sigma
_{u}^{+})$ and He$_{2}^{\ast }(a^{3}\Sigma _{u}^{+})$ dimers in
the liquid which decaying produce UV scintillation. A number of
publications~\cite {Bubble1,Bubble2, Bubble3,Bubble4} addressed an
unusual phenomenon in liquid helium such as the formation of
cavities around free electrons and excited
atoms and molecules. The dynamics of the formations of cavities is fast, $%
5\times 10^{-11}$ s \cite{Bubble1}, so that excited molecules do not decay
significantly during the time of cavity formation in which case the liquid
effects on the spectral properties are greatly reduced.

The luminescent spectrum of liquid and solid neon was observed by~%
\citet{Packard} in the range 700-900 \AA . It is interesting to note that
the solid has a narrow peak at 743 \AA \thinspace , but the liquid does not;
similar peaks in solid and liquid exist at 774 \AA\ with widths about 30 \AA %
. The luminescence of neon gas was also briefly investigated: at
low pressure $P<40$ Torr, a single narrow line was detected near
743 \AA . At higher pressures, the intensity of the 743 \AA\ line
decreased but the
second line at 800 \AA\ appeared. At low temperature, 25 K, the 800 \AA\ %
line\ does not persist.

Absorption and emission spectra of liquids can be obtained
approximately from the spectra of molecules. For example, in
Ref.\cite{Stockton} it was found that lines in the 600\AA\ band of
He$_{2}$ appear in the liquid emission spectrum and can be well
predicted from potential curves of the pair interaction using
Franck-Condon principle. We would like to apply this principle to
obtain theoretical UV emission in helium and neon liquids
(molecules) near 16 eV. The main peak of helium scintillation
according to Ref.\cite{Stockton} appears around 15.5 eV (800\AA ),
and neon liquid scintillation was observed at 16
eV~\cite{Packard}. For calculations of the wavelength of the
maximum of emission it is enough to know the equilibrium
interatomic distance ($R_{eq}$) of the lowest singlet excited
molecular state, the energy of this state, and the interatomic
energy of two ground state atoms at the distance $R_{eq}$.
Detailed spectral distribution can be obtained from the overlap of
wave functions of nuclear motion, according to the classical work
by~\citet{Condon}. Absorption transitions start from ground state
atoms which are separated by a relatively large distance. Thus the
small energy shift due to environment can be neglected and it is a
reasonable approximation to consider only the strongest allowed
atomic transitions which have transition energies close to the
emission energy of about 16 eV.

The helium interatomic or molecular potentials are
well-known~\cite {Sunil,Chabaloswski,Ginter,Komasa}. The potential
curves of the lowest excited states of Ne$_{2}$ are given in
Ref.\cite{Giessl}. The ground state of neon diatomic molecule is
investigated in~\cite{Neon1,Neon2,Cohen,Gilbert}. Analytical
expansions of the ground-state interatomic potentials of noble-gas
atoms that reproduce well the available vibrational spectra are
given in \cite{Neon3}. Most studies do not consider small
interatomic distances in detail which are important for our
application, so we supplement them with our own calculations
described below.

Our \textit{ab initio} ground-state interatomic potentials at
small distances are calculated using Pauli's principle. When two
closed-shell atoms have wavefunctions overlapping, electrons are
promoted to excited atomic states increasing the energy of each
atom. In helium, the repopulation increases the energy  by 22 eV
times overlap and in neon 16 eV times overlap. Wavefunction
overlaps are obtained using Dirac-Hartree-Fock potentials,
1s$^{2}$ in the case of helium and 1s$^{2}$2s$^{2}$2p$^{6}$ in the
case of neon. Overlaps of all electrons of one atom with those of
the other are taken into account. In the range of interatomic
energies of interest for self-absorption other effects such as
electrostatic direct and exchange interactions are weaker because
of screening. We have checked the magnitude of electrostatic
effects in case of helium, and found that the dominant
contribution comes from the Pauli principle.

\begin{figure}[tbp]
\centerline{\includegraphics*[scale=0.8]{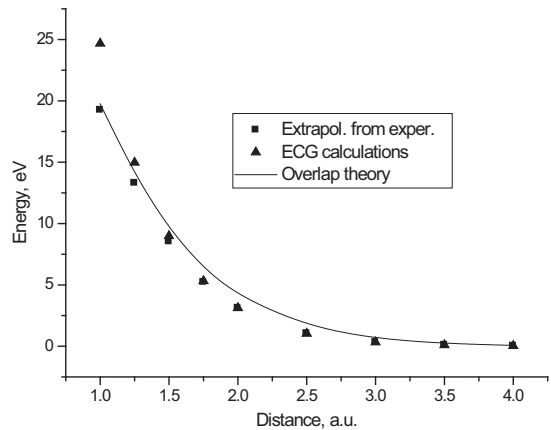}}
\caption{Comparison of the ground-state helium interatomic
potentials obtained from the overlap theory, the extrapolation
using the expansion of Ref.\protect\cite{Neon3}, and \textit{ab
initio} calculations with exponentially correlated Gaussian
functions \cite{Komasa}} \label{hemol}
\end{figure}
\begin{figure}[tbp]
\centerline{\includegraphics*[scale=0.8]{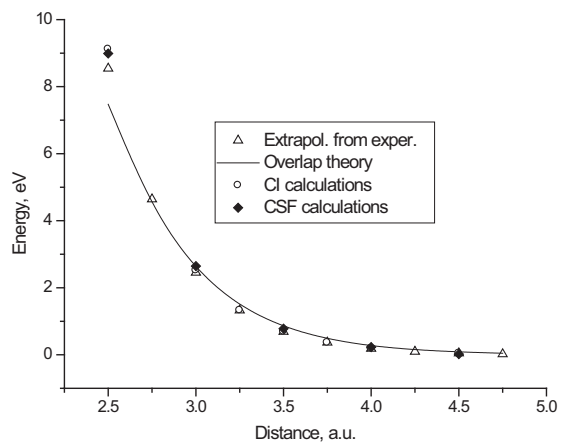}}
\caption{Comparison of the ground-state neon interatomic
potentials obtained from the overlap theory, the extrapolation
using the expansion of Ref.\protect\cite{Neon3}, with CI
method~\cite{Cohen}, and in SCF calculations~\cite{Gilbert}}
\label{nemol}
\end{figure}
In Fig.\ref{hemol} we show the comparison of the ground-state
helium interatomic potentials obtained from the overlap theory,
the extrapolation using the expansion of Ref.~\cite{Neon3}, and
\textit{ab initio} calculations with exponentially correlated
Gaussian (ECG) functions \cite{Komasa}. The expansion of
\cite{Neon3} agrees well with calculations of \cite{Komasa} up to
10 eV indicating good precision of both methods. The overlap
calculations agree with the expansion results over a larger range,
with some deviation in the middle of the curve due to
approximations of our theory. We can conclude that our method
describes well the repulsive interaction in helium and the
expansion method is suitable for calculations below 10 eV. In the
case of neon (Fig.\ref{nemol}), we find close agreement between
different methods: the overlap theory, the extrapolation from
experiment\cite{Neon3}, configuration-interaction (CI)
method~\cite{Cohen}, and self-consistent field (SCF) calculations
~\cite{Gilbert} below 8 eV. We also found agreement between our
method and that of Ref.\cite{Neon3} in argon, although not as
close as in neon. Therefore, the expansion of Ref.~\cite{Neon3}
gives the energies of ground state noble-gas dimers with
sufficient precision for calculations of emission spectra and the
analysis of self-absorption.
\begin{figure}
\centerline{\includegraphics*[scale=0.8]{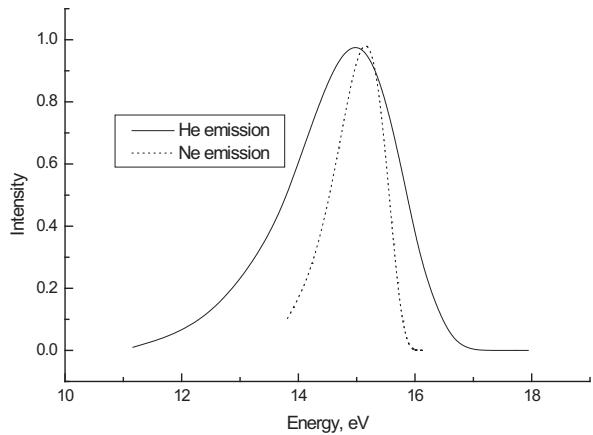}}
\caption{Theoretical helium and neon VUV emission spectra. The
intensity is normalized to the maximum} \label{heemit}
\end{figure}

The emission spectra of helium and neon molecules are shown in
Fig.\ref{heemit}. The initial He excited state has equilibrium
distance $R_{e}=1$ \AA, energy 18.2 eV and the harmonic-oscillator
frequency $\omega _{e}=1861$ cm$^{-1}$ \cite{Sunil}. Corresponding
values for the Ne $3s^3\Sigma_u$ initial state are: 1.75 \AA,
16.11 eV, and 540 cm$^{-1}$ \cite{Giessl,Cohen}. Neglecting
anharmonic corrections, we can write the nuclear wavefunction of
the lowest vibrational state as
\begin{equation}
\Psi (R)=\frac{\alpha }{\pi ^{1/2}}\exp [{-\frac{\alpha ^{2}(R-R_{e})^{2}}{2}%
}]
\end{equation}
here $\alpha =(2\pi \mu c\omega_e /\hbar )^{1/2}$, $\mu $ is the
reduced mass, $c$ is the speed of light. Although excited
vibrational states are always populated in the process of any
non-radiative excitation, very fast collisions and quenching will
transfer much of their population into lowest vibrational states.
Since only direct transitions are the most probable, the
absorption spectrum will have the weighting factor $|\Psi
(R)|^{2}$. The helium theoretical molecular spectrum is in better
agreement with the emission spectrum observed in condensed helium
discarge~\cite{Stockton,Huffman} than with that observed in liquid
\cite{Stockton,Surko}. High-pressure discharge emission has
maximum located at 15.3 eV \cite{Stockton,Huffman}, while
theoretical prediction is 15 eV, and the maximum of the emission
of liquid is located at 15.5 eV \cite{Stockton} and 15.8 eV
\cite{Surko}. The width at half maximum of discharge spectrum is 2
eV ~\cite{Stockton}, the same as our prediction, while the liquid
spectrum has the width about 3 eV \cite{Stockton,Surko}. From this
comparison, we can conclude that the liquid effects shift the
energy by 0.5-0.8 eV and broaden the spectrum by 1 eV. A similar
situation occurs in neon. High-pressure neon discharge gives the
maximum of emission at 15.5 eV, while the liquid has the maximum
at 16 eV \cite{Packard}; the theoretical prediction is 15.2 eV.
The theoretical width is about 1 eV, and width of the liquid
spectrum obtained with 1 eV resolution is about 2 eV. The liquid
effects result in shift of at least 0.5 eV and broadening less
than 1 eV. The shift is not easy to explain; we can suggest that
the interactions of more than two atoms can be responsible and the
distance between an excited atom and closest ground state atoms
can be different than that in the diatomic molecule.

Apart from calculations of emission spectra, we also investigated
various absorption mechanisms and calculated self-absorption
coefficients. In liquids as well as in dense gases in the vicinity
of an atomic or molecular resonance collisional broadening is very
important, and the absorption due to collisionally broadened
Lorentzian wings is expected to be very strong; however, the
impact approximation from which the Lorentzian shape-factor can be
derived is not valid for far-off-resonance transitions, which is
the case in our system, and instead more appropriate is to use
quasi-static approximation (see for example~\cite {sobelman}, pp.
241-245) in which the absorption profile is determined from
probabilities $W(R)dR$ of nearest particles within the range of
distances (R, R+dR) from a given atom. Either the classical
distribution or quantum-mechanical wavefunctions can be used to
obtain these probabilities. In the classical case, the Boltzman
distribution for low temperatures of the liquids gives very strong
exponential suppression for the detuning of order 1 eV and the
absorption can be neglected. On the other hand, the consideration
of wavefunctions of nuclear motion (the tunnelling into
classically-forbidden region) gives negligible increase in
absorption because the tunnelling probability is small and the
energy deficit will remain whether the tunnelling occurs in the
ground state, excited state, or both. Finally, we conclude that
the dominant absorption is likely due to wings of the
radiatively-broadened profile.

For large detuning and radiative broadening, the absorption
coefficient $\alpha _{a}$ is (see for example Ref.\cite{sat:b1})
\begin{equation}
\alpha _{a}=\frac{Ne^{2}\omega _{0}}{cm\epsilon _{0}}\frac{\gamma \omega f%
}{(\omega _{0}^{2}-\omega ^{2})^{2}+\gamma ^{2}\omega ^{2}}
\end{equation}
where $\gamma $ is the width of the resonance, $\omega _{0}$ is
the resonance frequency, $\omega $ is the frequency of the light,
$N$ is the number density,  $c$ is the speed of light, $\epsilon
_{0}$ is permittivity constant, $f$ is the oscillator strength of
the dominant atomic transition. The equation assumes CI units.
Note that the frequency dependence approaches Lorentzian for
detuning much smaller than the resonance frequency. The
self-absorption coefficients for liquid helium and neon calculated
using the above equation are given in Table~\ref{table1}.

The scattering lengths can be found from the theory of Rayleigh
scattering given by \citet{Landau1}:
\begin{equation}
\alpha_s=\frac{\omega^4}{6\pi c^4}kT\rho^2\kappa_T\left(\frac{\partial \epsilon}{%
\partial \rho}\right)_T^2
\end{equation}
We neglected the temperature derivative term which was estimated
to be small in Ref.\cite{Seidel}. The compressibility $\kappa_T$
of He and Ne are 208$\times$10$^{-10}$ and 4.95$\times$10$^{-10}$
cm$^2$/dyne; $\rho$ is the density. The dielectric constant is
related to the refractive index: $\epsilon=n^2$. The refractive
index can be calculated using the equation (see for example
Ref.~\cite{Smith})
\begin{equation}
\frac{n^{2}-1}{n^{2}+2}=\frac{\alpha ^{2}Na_{0}\lambda ^{2}}{3\pi }\sum_{i}%
\frac{\lambda _{i}^{2}f_{i}}{\lambda ^{2}-\lambda _{i}^{2}}
\end{equation}
where $\alpha $ is the fine-structure constant, $N$ is the
number-density of atoms, and $a_{0}$ is the Bohr radius. The
summation is extended over continuum. For neon, we included 9
values from Ref.\cite{neonepr} and estimated remaining
contribution from the Thomas-Reiche-Kuhn rule. For helium we
calculated 40 oscillator strengths with CI method using B-spline
basis sets obtained in a cavity 30 a.u. The continuum contribution
in helium is larger than that in neon because of larger detuning
from the dominant transition for the wavelengths of interest. To
check our continuum contributions in helium, we calculated the
refraction index at wavelengths in the range from 168 to 288 nm to
compare with available the precise measurement by~\citet{Smith}.
We found that our ab initio values agree well with the
experimental values. The largest uncertainty in our helium
calculations at 78 nm comes from the replacement of continuum
states with quasicontinuum states, and in neon calculations from
the estimation of continuum contributions using the sum rule. The
results of calculations of scattering coefficients are given in
Table~\ref{table1}.
\begin{table}
\caption{Liquid helium and neon absorption and scattering
coefficients. Parenthesis denote of the uncertainty of
calculations in the last digit. } \label{table1}
\begin{tabular}{lcccccc}
\hline
Liquid & $n,cm^{-3}$ & $\lambda ,nm$ & $\lambda _{0},nm$ & $\alpha _{a},$ m%
$^{-1}$ & $\frac{n^{2}-1}{n^{2}+2}$ & $\alpha _{s},$ m$^{-1}$ \\
\hline
He & 1.88$\times10^{22}$ & 78 & 58.43 & 0.36 & 0.028(2) & 0.21(4) \\
Ne & 3.60$\times10^{22}$ & 80 & 73.59 & 2.02 & 0.15(4) & 1.5(8) \\
\hline
\end{tabular}
\end{table}

We found that our results for scattering lengths and dielectric
constants, $\epsilon_{He}=1.086\pm{0.006}$ and
$\epsilon_{Ne}=1.5\pm{0.2}$, are in agreement with those obtained
by~\citet{Seidel} using extrapolation of experimental refraction
indices: $\epsilon_{He}=1.077$, $\epsilon_{Ne}=1.52$, $R_{He}=600$
cm , and $R_{Ne}=60$ cm. Our calculations has been done to check
the previous results and to evaluate the uncertainty.

In our calculations we found that absorption lengths are smaller
than the size of the neutrino detector, so that it is important to
know whether the excited atom re-emits an absorbed photon or it is
quenched non-radiatively. Because for non-radiative decay triplet
and singlet states are similar and in Ref.\cite{kopeliovich} it
has been found that lifetime of the helium triplet state in solid
exceeds 15 s, we can conclude that quenching is a slow process and
the total emission power will not be reduced much. The
absorption-re-emission process is equivalent to the scattering in
this case.

In this paper we have analyzed UV emission, self-absorption, and
scattering of liquid helium and neon. For emission spectra we have
found reliable data for helium and neon molecules and developed
our own model of repulsive interaction to check and understand the
existing calculations. Using potential curves we obtained emission
spectra which agree well with experimentally observed condensed
discharge spectra, although liquid emission is shifted due to
liquid effects. We have considered various mechanisms of
self-absorption and found that the radiative broadening is the
dominant mechanism if the emission absorption offset exceeds 1 eV.
The self-absorption is significant at the scale of the proposed
neutrino detector and can impose restrictions on the experiment;
fortunately, non-radiative relaxation seems to be slow and the
absorption results mostly in re-emission so that it is similar to
scattering in terms of conservation of the number of photons, but
dissimilar to it in terms of photons' propagation dynamics. We
have also calculated dielectric constants and Rayleigh scattering
parameters. Results are in agreement with previous calculations.
We understand now dominant effects of the emission,
self-absorption, and scattering, but some questions remain and
further investigation is necessary: the study of the liquid shifts
and other possible liquid effects, the study of quenching
mechanisms, the measurement of scattering and absorption lengths.
The author is grateful to Dr. McKinsey for valuable discussions
and comments on the manuscript.

\end{document}